\documentclass[aps,prl,superscriptaddress,reprint]{revtex4-2}

\newcommand{\Date}{June 21, 2026}

\newcommand{\se}[1]{\paragraph{#1.---}}

\newcommand{\pa}[1]{\paragraph{#1}}

\newcommand{\bfthm}[1]{\noindent\textbf{#1}}

\usepackage{amsmath}
\usepackage{amssymb}
\usepackage{amsfonts}

\newcommand*{\dd}{\mathop{}\!\mathrm{d}}

\DeclareMathOperator{\Span}{span}

\DeclareMathOperator{\Arcsin}{Arcsin}

\usepackage{amsthm}
\theoremstyle{definition}
\newtheorem{theorem}{Theorem}
\newtheorem*{theorem*}     {定理}
\newtheorem{lemma}
{Lemma}
\newtheorem*{lemma*}       {補題}
\newtheorem{proposition}
{Proposition}
\newtheorem*{proposition*}{命題}
\newtheorem{corollary}
{Corollary}
\newtheorem*{corollary*}{系}
{Definition}
\newtheorem*{definition*}
{Definition}
{Conjecture}
\newtheorem*{conjecture*}
{Conjecture}
{Assumption}
\newtheorem*{assumption*}
{Assumption}

\newcommand{\comment}[1]{}

\newtheorem*{remark*}      {Remark}

\newcommand{\abs}[1]{{\left\vert #1 \right\vert}}
\newcommand{\spp}[1]{^{(#1)}}

\newcommand{\nabs}[1]{{| #1 |}}
\newcommand{\bigabs}[1]{{\bigl | #1 \bigr |}}
\newcommand{\Bigabs}[1]{{\Bigl | #1 \Bigr |}}

\newcommand{\paren}[1]{{\left( #1 \right)}}

\newcommand{\nparen}[1]{{( #1 )}}
\newcommand{\bigparen}[1]{\bigl({#1}\bigr)}
\newcommand{\Bigparen}[1]{\Bigl({#1}\Bigr)}

\newcommand{\brac}[1]{{\left\{ #1 \right\}}}

\newcommand{\bigbrac}[1]{\bigl\{{#1}\bigr\}}

\newcommand{\brak}[1]{{\left[ #1 \right]}}
\newcommand{\bigbrak}[1]{\bigl[{#1}\bigr]}

\newcommand{\tht}{\theta}

\newcommand{\pder}[2]{{\partial{#1}\over\partial{#2}}}

\newfont{\bg}{cmr10 scaled\magstep4}                    
\newcommand{\bigzerou}{\smash{\lower1.7ex\hbox{\bg 0}}}

\newcommand{\nn}{\notag \\ }

\newcommand{\di}{\displaystyle}
\newcommand{\ma}[1]{{$\di #1$}}

\newcommand{\HH}{{\cal H}}

\newcommand{\II}{{\cal I}}

\newcommand{\CC}{{\cal C}}

\newcommand{\TT}{{\cal T}}

\newcommand{\crl}[1]{[-\infty,\infty]}
\newcommand{\ba}[1]{\bar{#1}}

\newcommand{\ket}[1]{\left|{#1}\right\rangle}

\newcommand{\bra}[1]{\langle{#1}|}
\newcommand{\sw}[2]{\bra{#1}{#2}\ket{#1}}

\newcommand{\ti}{\tilde}
\newcommand{\ha}{\hat}
\newcommand{\wh}{\widehat}

\newcommand{\da}[1]{#1^\dag}

\newcommand{\ta}{\tau}

\newcommand{\acom}[2]{\left\{{#1},{#2}\right\}}
\newcommand{\ac}{\acom}

\renewcommand{\a}{\alpha}
\renewcommand{\b}{\beta}
\renewcommand{\c}{\gamma}
\renewcommand{\d}{\delta}
\newcommand{\De}{\varDelta}

\newcommand{\ep}{\varepsilon}

\newcommand{\Gm}{\Gamma}

\renewcommand{\l}{\lambda}

\newcommand{\La}{\varLambda}

\newcommand{\m}{\mu}

\newcommand{\om}{\omega}
\newcommand{\Om}{\Omega}

\renewcommand{\r}{\rho}

\newcommand{\f}{\frac}

\newcommand{\q}{\quad}
\newcommand{\qq}{\qquad}

\newcommand{\ve}[1]{\boldsymbol{#1}}
\newcommand{\we}{\wedge}

\newcommand{\norm}[1]{\left\|#1\right\|}

\newcommand{\bignormop}[1]{\bigl\|#1\bigr\|_{\mathrm{op}}}

\newcommand{\nnormop}[1]{\|#1\|\sbt{op}}
\newcommand{\norminf}[1]{\left\|#1\right\|_\infty}
\newcommand{\normi}[1]{\left\|#1\right\|_\infty}

\renewcommand{\ge}{\geqslant}
\renewcommand{\geq}{\geqslant}
\renewcommand{\le}{\leqslant}
\renewcommand{\leq}{\leqslant}

\newcommand{\sbt}[1]{_{\text{\textrm{#1}}}}
\newcommand{\tx}{\text}

\newcommand{\dsn}[1]{ds^2_{N_\Gm}}

\newcommand{\toinf}{\to\infty}

\newcommand{\intinf}{\int_{-\infty}^\infty}

\newcounter{ssetmp}

\makeatletter 
\def\overbracket#1{\mathop{\vbox{\ialign{##\crcr\noalign{\kern3\p@} 
\downbracketfill\crcr\noalign{\kern3\p@\nointerlineskip} 
$\hfil\displaystyle{#1}\hfil$\crcr}}}\limits} 
\def\underbracket#1{\mathop{\vtop{\ialign{##\crcr 
$\hfil\displaystyle{#1}\hfil$\crcr\noalign{\kern3\p@\nointerlineskip} 
\upbracketfill\crcr\noalign{\kern3\p@}}}}\limits} 
\def\overparenthesis#1{\mathop{\vbox{\ialign{##\crcr\noalign{\kern3\p@} 
\downparenthfill\crcr\noalign{\kern3\p@\nointerlineskip} 
$\hfil\displaystyle{#1}\hfil$\crcr}}}\limits} 
\def\underparenthesis#1{\mathop{\vtop{\ialign{##\crcr 
$\hfil\displaystyle{#1}\hfil$\crcr\noalign{\kern3\p@\nointerlineskip} 
\upparenthfill\crcr\noalign{\kern3\p@}}}}\limits} 
\def\downparenthfill{$\m@th\braceld\leaders\vrule\hfill\bracerd$} 
\def\upparenthfill{$\m@th\bracelu\leaders\vrule\hfill\braceru$} 
\def\upbracketfill{$\m@th\makesm@sh{\llap{\vrule\@height3\p@\@width.7\p@}}%
\leaders\vrule\@height.7\p@\hfill 
\makesm@sh{\rlap{\vrule\@height3\p@\@width.7\p@}}$} 
\def\downbracketfill{$\m@th 
\makesm@sh{\llap{\vrule\@height.7\p@\@depth2.3\p@\@width.7\p@}}%
\leaders\vrule\@height.7\p@\hfill 
\makesm@sh{\rlap{\vrule\@height.7\p@\@depth2.3\p@\@width.7\p@}}$} 
\makeatother 

\newcommand{\eqb}{_{\text{eq}}}
\newcommand{\neqb}{_\mathrm{neq}}
\newcommand{\nqb}{\neqb}

\newcommand{\ceil}[1]{\left\lceil#1\right\rceil}

\newcommand{\Bigceil}[1]{\Bigl\lceil#1\Bigr\rceil}

\newcounter{enumicont}
{%
  \setcounter{enumicont}{\value{enumi}}
  \begin{enumerate}%
  \setcounter{enumi}{\value{enumicont}}
}%
{\end{enumerate}}

{%
  \pagebreak
  \setcounter{page}{1}
}%
{\pagebreak}
{\pa{#1}}%
{\hfill {\red ■}}

{\begin{framed}  \begin{lemma}}%
{\end{lemma}\end{framed}}

{\begin{framed}  \begin{proposition}}%
{\end{proposition}\end{framed}}

{\begin{framed}  \begin{theorem}}%
{\end{theorem}\end{framed}}

{\begin{framed}  \begin{corollary}}%
{\end{corollary}\end{framed}}

\newcommand{\ketvac}{\ket{\tx{vac}}}

\newcommand{\xx}{{\ve x}}
\newcommand{\yy}{{\ve y}}
\renewcommand{\aa}{{\ve \a}}
\newcommand{\bb}{{\ve \b}}
\newcommand{\cc}{{\ve \c}}
\newcommand{\mm}{{\ve m}}
\newcommand{\zz}{{\ve 0}}
\newcommand{\eql}{equilibrium}

\newcommand{\nql}{nonequilibrium}
\newcommand{\noneq}{nonequilibrium}
\newcommand{\bo}[1]{{\cal B}(#1)}
\newcommand{\bd}{\delta}

\newcommand{\timescale}{timescale}

\begin{document}
\title{%
      	Timescale for macroscopic equilibration in isolated quantum systems:
        \\
        a rigorous derivation for free fermions 
}
\author{Takashi Hara}
\email{hara@math.kyushu-u.ac.jp}
\affiliation{Faculty of Mathematics, Kyushu University, 
	Fukuoka 819-0395, Japan}
\author{Tatsuhiko Koike}
\email{koike@phys.keio.ac.jp}
\affiliation{Department of Physics, Keio University,
  Yokohama 223-8522, Japan}
\date{\Date}

\begin{abstract}
  For a 
  class of translation-invariant free-fermion systems 
(including those with uniform nearest neighbor hopping) 
on a $d$-dimensional $L \times \cdots \times L$ hypercubic lattice, 
we prove that, starting from an arbitrary pure initial state, 
the system equilibrates with respect to 
the coarse-grained density within a {\timescale} of order $L$.  
This scaling is optimal, 
since there exist initial states whose equilibration requires time of order $L$. 
Our result establishes $O(L)$ as the equilibration {\timescale}, 
as is expected in normal macroscopic systems with a conserved quantity, 
such as total number of particles.
\end{abstract}
\maketitle
\thispagestyle{empty}

\se{1. Introduction}

Nearly a century ago, 
von Neumann~\cite{neumann1929beweis,*von2010proof}
presented an idea concerning
the foundations of quantum statistical mechanics: 
Any initial state of an isolated quantum system 
evolves into an equilibrium, so that  quantum dynamics alone, 
without  any
\emph{ad hoc} assumptions, can explain thermalization
(see also \cite{GLTZ-2010,Reimann2015-PhysRevLett.115.010403}).  
Although his idea had somewhat been forgotten 
for a long time,
researches in this direction, 
which show thermalization or equilibration~\footnote{%
A system is said to equilibrate for an observable, if the expectation
value of the observable approaches 
a stationary value.  A system thermalizes, if it equilibrates for
almost all (macroscopic) observables and 
their stationary values are given by a Gibbs ensemble. 
We in this Letter will consider equilibration.
},
are now flourishing.
In particular, it has been demonstrated
that a pure initial state evolves into an equilibrium
state under quantum dynamics in some general or concrete settings 
\cite{Tasaki1998-PhysRevLett.80.1373,Rigol2008Nature,Reimann2008-PhysRevLett.101.190403,Goldstein-Lebowitz-2010-PhysRevE.81.011109,Reimann_2012isol,ShiraishiTasaki2024JSP,RoosSTT2024}.  
Tasaki~\cite{Tasaki2024freeexpansion}
recently showed in  a nondegenerate free fermion chain 
that 
\emph{macroscopic} density observables 
equilibrate for \emph{any} initial state.

It should be noted, however, that  the issue of the
\emph{{\timescale} required for macroscopic 
equilibration has not been well addressed}.
Statements about equilibration are usually 
proved for ``sufficiently long (but finite) time''
(see, e.g.,
Refs.~\cite{Tasaki1998-PhysRevLett.80.1373,Rigol2008Nature,Reimann2008-PhysRevLett.101.190403,Goldstein-Lebowitz-2010-PhysRevE.81.011109,Reimann_2012isol,ShiraishiTasaki2024JSP,RoosSTT2024,Tasaki2024freeexpansion}). 
[The situation is different
for equilibration of local observables. 
See Section 1.1.]
Although there are several works
\cite{GHT2013-PhysRevLett.111.140401,GHT2015,Malabarba-Linden-Short-2014-PhysRevE.90.012121} 
which provide close to optimal {\timescale} of equilibration 
in abstract settings 
\footnote{
  Von Neumann's original idea turned out to be
  too optimistic in the sense that it
  leads to an unrealistically short {\timescale}
  of equilibration~\cite{GHT2015}.  
},  
there are hardly any good estimates 
for {\timescale}s of equilibration 
in concrete physically realistic short-ranged systems. 

In this Letter, 
we address this problem of {\timescale}, 
and obtain optimal {\timescale} of macroscopic equilibration for a system of free fermions. 
More precisely, we consider $N$ spinless
fermions  with uniform nearest-neighbor hopping on a $d$-dimensional lattice $L^d$ 
(of linear size $L$, with periodic boundary condition), and its 
macroscopic density observable $\hat\rho$.  
We look into the time evolution of the system from an initial state, and 
focus on the fraction of time duration during which expectation of the density
$\hat\rho$ macroscopically deviates from its expected average value, $N/L^d$.  
We prove that the fraction goes to zero in the thermodynamic limit 
(i.e.  $L \to \infty$ with $N/L^d$ fixed), for \emph{every} initial state 
and for any $\tau \geq L \times g(L)$, where $g(L)$ is an arbitrary
function that diverges as $L \to \infty$.   
Since there exist initial states whose equilibration requires time of order $L$, 
our result establishes $O(L)$ as the macroscopic equilibration {\timescale}. 

To the best of our knowledge, this is the first time that 
\emph{a realistic {\timescale} of macroscopic equilibration has been rigorously established} for
isolated physically natural quantum systems.

\se{1.1. Comparison with works on systems with many conserved quantities}

For systems with many conserved quantities
(including free fermion systems considered in this Letter), 
equilibration for various {local} observables 
has been studied in
detail~\cite{Bartell-PhysRevLett.100,Cramer-Eisert-2010,Cazalilla-PhysRevE.85.011133,Sotiriadis_2014,Gluza-et-al-PRL117,Vidmar-Rigol2016JSM,Gluza-Eisert-Farrelly2019,Murthy-Srednicki-PRE100,Gluza-Eisert-Farrelly2019}: 
If the initial state and the dynamics satisfy certain conditions,
various local observables equilibrate  rather rapidly, and 
this local equilibrium persists up to 
a timescale of the system size $L$. 
On the other hand,
results on \emph{macroscopic} observables seem to be scarce, 
and it is not clear what happens
on larger timescales. 
Although physical intuition suggests that the system macroscopically
equilibrates
after a timescale of order $L$, and there
are some works 
which
partially support this (see, e.g., 
\cite[eq.~(51)]{Murthy-Srednicki-PRE100}),
it is not an easy task to establish this fact.

This work is  complementary to 
existing works: we focus on a \emph{macroscopic} observable, 
i.e.,  macroscopic density, 
and investigate its behavior 
\emph{beyond the time scale of the system size} $L$.
We rigorously
establish that the time average of the macroscopic density approaches
and remains in 
its equilibrium value after the timescale of order $L$, 
for any initial state. 

One more remark is in order.  
Since our system is integrable, it is expected to approach 
equilibrium states described by generalized Gibbs
ensembles. 
In this sense, 
our system never thermalizes~\cite{Note1}. 
However,  
we hope our analysis would shed some light on understanding 
thermalization in non-integrable systems as well.


\se{2. The setup and the main result} 

We now explain our basic setup in more detail and state our main result. 
\se{2.1. Free fermions on a lattice} 
We use the unit $\hbar=1$. 
We consider a system of free fermions on a $d$-dimensional
hypercubic lattice $\La$ of size $L^d=:V$, 
with periodic boundary conditions.
We assume $L$ is odd,  
while we can also easily deal with even $L$'s by 
a slight change of  notation. 
The sites of $\La$ are labeled by
$\xx=(x^1, \dots,x^d) \in {\mathbb Z}^d$, $\abs{x^\m}<L/2$. 
We fix the density $\ba\r>0$, 
and choose the total number $N$ of 
fermions as the largest integer such that $N/V \le \ba\r$.  
The Hamiltonian is given by 
\footnote{For simplicity, we here treat the simplest model, namely,
  the one with uniform nearest-neighbor hopping. 
Our result should extend to models in 
which $\abs{\Om_\mm(E)}$ 
of \eqref{eq-Omega-def} 
is sufficiently regular.
}
\begin{align}
	\ha H:= \sum_{\norm{\xx-\yy}=1}\da {\ha c}_\xx \ha c_\yy, 
	\qquad \|\xx\| := \Bigl ( \sum_{\mu=1}^{d} (x^\mu)^2 \Bigr )^{1/2}
	, 
  	\label{eq-ham}
\end{align}
where $\da c_\xx$ and $c_\xx$ 
are spinless fermion's creation and annihilation operators at $\xx\in\La$,
which satisfy the canonical anticommutation relations  
$
  \ac {\ha c_\xx}{\ha c_\yy}=  \ac {\da{\ha c}_\xx}{\da{\ha c}_\yy}=0,\, 
  \ac {\ha c_\xx}{\da{\ha c}_\yy}=\d_{\xx,\yy}. 
$
The creation operator of momentum
eigenstate $\ket\aa$ of eigenvalue $2\pi\aa/L$,
where $\aa = (\alpha^1, \dots, \alpha^d)\in {\mathbb Z}^d$ and $\abs{\a^\m}<L/2$, is given by 
\begin{equation}
  	\da {\ha a}_\aa
  	:=\f1{\sqrt V}\sum_\xx e^{2\pi i\aa\cdot\xx/L}
  	\da {\ha c}_\xx
	. 
   	\label{eq-a}
\end{equation}
The $N$-particle Hilbert space $\HH$ is spanned
by $\da {\ha a}_{\aa_1} \cdots \da {\ha a}_{\aa_N}\ketvac$, 
where $\aa_i \neq \aa_j$ for $i \neq j$, and $\aa$'s are 
canonically ordered. 
The Hamiltonian \eqref{eq-ham} can be written as
\begin{equation}
  	\ha H=\sum_\aa E_\aa \da{\ha a}_\aa {\ha a}_\aa,
  	\q
  	E_\aa:=\sum_{\m=1}^d
  	2\cos\f{2\pi \a^\m}L. 
  	\label{eq-H-in-a}
\end{equation}

Our macroscopic observable is 
the number density 
\begin{equation}
  	\ha\r_{\ve c}:=
  	\f1{l^d}
  	\sum_{\xx\in \bo{\ve c}}\da {\ha c}_\xx \ha c_\xx
  	\label{eq-rho}
\end{equation}
of fermions in a macroscopic box centered at $\ve c\in\La$ and
of volume $l^d$: 
$
\bo{\ve c}:=\brac{\xx\in\La; \normi{\xx-\ve c}<\f l2} 
$. 
Here $\normi\xx:=\max\brac{|x^1|, \dots, |x^d|}$. 
To simplify the notation, we have assumed that $l$ is odd. 
We prepare $n^d$ boxes to cover the whole lattice, where
$n:=\ceil{\tfrac{L}{l}}$
is the least integer greater than or equal to $\tfrac{L}{l}$. 
The set of centers of the boxes is 
$\CC=\bigl\{jl; -\f{n}2<j\le \f{n}2 , \, j \in \mathbb{Z} \bigr\}^d$,
where 
the power $d$ denotes the $d$-fold direct set product. 
In what follows, $n$ can be large but still $O(1)$, 
while  $L$ and $l$ are macroscopically large. 

\se{2.2. Nonequilibrium subspace} 
We decompose the Hilbert space $\HH$ into 
{\eql} subspace $\HH\eqb$ and {\nql} subspace $\HH\neqb$ as follows. 
Since all $\ha\r_{\ve c}$ mutually commute, 
there is an orthonormal basis $\brac{\ket{e_ j}}$ of $\HH$
such that each 
$\ket{e_ j}$ is a simultaneous eigenstate of all $\ha\r_{\ve c}$ ($\ve c\in\CC$). 
Let us introduce an operator 
$\wh{\De\r}_{\ve c}:=\ha\r_{\ve c}-N/V$, 
the deviation of density from the mean density, and 
let $\ket{e_ j}$ be an eigenvector: 
$\wh{\De\r}_{\ve c}
\ket{e_ j}=\l_{j,\ve c}\ket{e_ j}$. 
Define 
\begin{align}
  	\HH\eqb:=\Span
  	\brac{\ket{e_j}; \abs{ \l_{j,\ve c} }\le \ep \ba \r 
  	\tx{ for all } j\tx{ and } \ve c},
\end{align}
with a small constant $\ep>0$ of our choice 
($\ep=10^{-10}$, for example). 
Define $\HH\neqb:=\HH\eqb^\perp$, 
the orthogonal complement of $\HH\eqb$ in $\HH$. 
When we measure macroscopic observables $\ha\r_{\ve c}$ 
($\ve c\in\CC$) 
on a state $\ket\Psi\in\HH\eqb$, 
all measurement outcomes are close ($\le\ep\ba\r$) to $N/V$; 
so we necessarily judge that the system is in equilibrium. 
On the other hand, when we measure $\ha\r_{\ve c}$ on 
$\ket\Psi\in\HH\nqb$, 
the outcome for some $\ha\r_{\ve c}$ may be 
far ($>\ep\ba\r$) from $N/V$; so 
we may find that the system is out of equilibrium. 
Thus, it is natural to introduce the projection operator 
$\ha P\nqb$ onto the subspace $\HH\nqb$ to extract the
``{\noneq} component'' of any state $\ket\Psi\in\HH$.

\se{2.3. Main Result}
Consider the time evolution
$\ket{\Psi(t)}$ from an 
\emph{arbitrary} initial state $\ket{\Psi(0)}\in\HH$. 
Let   
\begin{align}
	\bd_a(\tau,L):=
  	\Bigparen{ \f{\log(\tau/L)}{\tau/L} }^a
  	+ \Bigparen{ \f{(\log L)^{2d}}L }^a
  	\label{eq-bd}
\end{align}
for $a>0$.
This is small when $\tau/L$ and $L$ are large,
and, in particular,
$\bd_{a}(\tau,L)\to0$ when $\tau/L$ and $L$ go to infinity.

Define the set of ``{\noneq} time'' during time interval $[t_1, t_2]$ as 
\begin{align}
  	\TT^{[t_1, t_2]}_{\delta, \,\ket{\Psi(0)}}
  	:=
  	\bigl \{ t\in[t_1, t_2];
 	\bra{\Psi(t)}\ha P\nqb\ket{\Psi(t)} >  \delta \bigr \}
\end{align}
and let $\bigl | \TT^{[t_1, t_2]}_{\delta, \,\ket{\Psi(0)}} \, \bigr |$ 
be the total duration (i.e., the Lebesgue measure) of $ \TT^{[t_1, t_2]}_{\delta, \,\ket{\Psi(0)}}$. 
Our main result is the following.

\smallskip

\bfthm{Theorem~1.}
The system equilibrates on the {\timescale} of order $L$. 
More precisely, let $\bar\tau(L) = L \times g(L)$, where $g(L)$ is an
arbitrary function that 
diverges as $L \to \infty$.   
Then, for any $\tau > \bar\tau(L)$ and for \emph{every} $N$-fermion initial state $\ket{\Psi(0)}$,
the fraction of {\noneq} time during the interval $[0, \tau]$ satisfies,
for $\delta = \delta_{\!\frac{1}{2}}\!(\tau, L)$, 
\begin{align}
	\frac{1}{\tau} \, \bigl | \TT^{[0, \tau]}_{\delta, \,\ket{\Psi(0)}} \, \bigr | \to 0 
\label{eq-ratio-noneq-time.0}
\end{align}
in the thermodynamic limit $L \to \infty$.

\medskip

Theorem~1 establishes $O(L)$ as the {\timescale} of equilibration 
for this system, because 
as the Lieb-Robinson bound \cite{LiebRobinson72} shows,
there are initial states whose
equilibration requires time of $O(L)$ \footnote{
  For example, when $\bar\rho$ is small, for an initial state where 
  the fermions are concentrated around the origin,
  there is an $L$-independent  constant $C>0$ such that 
  $\bigl | \sw{\Psi(t)}{\wh{\De\r}_{\ve c}} \bigr | <  \ba\r/2$ for
  $\abs t< C L$  and  $\ve c = (c,\dots,c)$,
  where $c$ is the largest integer such that $c\le L/3$. 
}. 
Note that the convergence of \eqref{eq-ratio-noneq-time.0} is uniform in $\ket{\Psi(0)}$; 
the speed of convergence 
does not depend on the choice of $\ket{\Psi(0)}$ for each $L$, 
as Theorem~1$'$ shows.

Theorem~1 is an immediate consequence of the 
following technical theorem, 
because $\bigl | \TT^{[0, \tau]}_{\delta, \,\ket{\Psi(0)}} \, \bigr | \leq  \bigl | \TT^{[-\tau, \tau]}_{\delta, \,\ket{\Psi(0)}} \, \bigr |$. 

\smallskip

\bfthm{Theorem~1$'$.}
Suppose $L > 10^4$ and $ \tau > 2L$.  
For \emph{every} $N$-fermion initial state $\ket{\Psi(0)}$,
the fraction of {\noneq} time during the interval $[-\tau, \tau]$ 
satisfies,
for $\delta = \delta_{\!\frac{1}{2}}\!(\tau, L)$, 
\begin{align}
	\frac{1}{2 \tau} \, \bigl | \TT^{[-\tau, \tau]}_{\delta, \,\ket{\Psi(0)}} \, \bigr |
  	\le
  	K_d \, \f{n^{3d}}{(\ep\ba\r)^2}\;\bd_{\f12}(\ta,L) , 
  	\label{eq-ratio-noneq-time}
\end{align}
where $K_d>0$ is a constant that depends only on $d$.

\smallskip

\se{Remarks}

(a) We take advantage of the \emph{macroscopic} nature of our observables
$\hat \rho_{\ve c}$,
where $n= \lceil \tfrac{L}{l} \rceil$ is $O(1)$,  in an essential way.
Remark (a) of Section 4.1 explains how the macroscopic nature is
utilized in our proof.

(b) We are \emph{not} assuming any  
kind of randomness in the system Hamiltonian. 
We are allowing degeneracy in the energy spectrum.

(c)
Although not explicitly stated, our proof does
work equally well for other systems, 
with the same result. 
For example, it works for nondegenerate Hamiltonians considered 
in \cite{Tasaki2024freeexpansion}.  The proof works for 
more general (e.g., not necessarily nearest-neighbor)
hopping of fermions for which an analogue of Lemma~5 holds.

(d)
For fixed $L$, the fraction
of the nonequilibrium time, i.e., the LHS of
\eqref{eq-ratio-noneq-time.0}, 
and the LHS of \eqref{eq-timeav-bd} below, do not go to
zero even when $\tau\to \infty$.  
They do vanish in the thermodynamic limit $L\toinf$ 
(while keeping $\ta/L>2$).

(e) 
Our system of free fermions never shows equilibration in 
momentum. That is, if we start from an eigenstate of momentum, the system stays in that 
state forever.  
It is of utmost importance to prove results similar to Theorem~1, 
for non-integrable 
systems that show equilibration \emph{both} in space and momentum
variables. 

\medskip

\se{3. Our key estimate for time average}
Theorem~1$'$ is proved by Proposition~2 below, 
which 
connects the {\noneq} time duration
$\bigl | \TT^{[-\tau, \tau]}_{\delta, \,\ket{\Psi(0)}} \, \bigr |$
with a time average of the
squared deviation $(\wh{\De\r}_{\ve c})^2$. 
We define our time average
$[\cdots]_\tau$
for a function of time $t$
by 
\begin{align}
  	[\cdots]_\ta:=\intinf \!\!dt \, f_\ta(t) (\cdots) ,
  	\q f_\ta(t):=\f1{\pi\ta}
	\Bigl ( \f{\sin(t/\ta)}{t/\ta} \Bigr )^2.
  	\label{eq-timeav-def}
\end{align}
The function $f_\ta(t)$ has significant values around the interval 
$[-\ta,\ta]$.   
The reason for our specific choice \eqref{eq-timeav-def} for $f_\tau$ will 
be clear later in Section~4.2. 
For this time average we have the following, which is our key estimate:

\smallskip

\bfthm{Proposition~2. }
Suppose 
$L>10^4$ and $\ta>2L$. 
Then, for \emph{every} $N$-fermion initial state $\ket{\Psi(0)}$ 
and for any $c \in {\cal C}$, 
we have 
\begin{align}
  	\brak{ \bigl \langle \Psi(t) \big | (\wh{\De\r}_{\ve c})^2 \, \big | \Psi(t) \bigr \rangle }_\ta
  	\le 
  	\f{K_d}3 
  	{n^{2d}}\;
  	\bd_1(\ta,L), 
  	\label{eq-timeav-bd}
\end{align}
where $K_d$ is the same constant as in Theorem~1$'$. 

\smallskip

We now show that Proposition~2 implies Theorem~1$'$ 
using the following two simple lemmas,
whose proofs are given in Appendices A and B. 

\smallskip

\bfthm{Lemma 3.}
The expectation values of
$\hat{P}\nqb$ and $ (\wh{\De\r}_{\ve c}){}^2$ are related by 
\begin{equation}
  	\sw\Psi {\hat{P}\nqb}
  	\le
  	\frac{1}{(\ep\ba\r)^2} \, 
  	\bigl \langle \Psi \big | \, 
  	{  \sum_{\ve c}  (\wh{\De\r}_{\ve c}){}^2}
  	\, \big | \Psi \bigr \rangle, 
\end{equation}
for any $\ket\Psi\in\HH$. 

\smallskip

\bfthm{Lemma 4.} (``Markov inequality'')
Suppose a function $X$ of time $t$ satisfies 
$X(t)\ge0$ and $\brak{X(t)}_\ta\le B$ with a 
constant $B>0$. 
Then, for any $t$-independent $X_1>0$, 
the fraction of the total time duration 
with $X(t)>X_1$ in the interval $[-\ta,\ta]$ 
is less than $3 B/X_1$. 

\smallskip

\se{Proof of Theorem~1$'$, based on Proposition~2}
This is a standard argument. 
By Proposition~2, Lemma~3, and $\abs \CC\le n^d$, 
we have 
$
  	\bigl [ \sw{\Psi(t)} {\hat{P}\nqb} \bigr ]_\ta
  	\le n^d/{(\ep\ba\r)^2}
  	\times
  	\paren{\tx{RHS of \eqref{eq-timeav-bd}}}
  	=:B. 
$
Then, by Lemma~4 with this $B$ 
and $X_1=\bd_{\f12}(\ta,L)=:\d$, we have
\begin{align}
	\frac{1}{2 \tau} \, \bigl | \TT^{[-\tau, \tau]}_{\delta,
  \,\ket{\Psi(0)}} \, \bigr | 
  	\le
  	3
  	\f{n^d}{(\ep\ba\r)^2}
  	\f{\paren{\tx{RHS of \eqref{eq-timeav-bd}}}}
  	{\bd_{\f12}(\ta,L)}. 
\end{align} 
Noting
$\bd_{1}(\ta,L)\le \nparen{ \bd_{\f12}(\ta,L) }^2$,
we obtain Theorem~1$'$. 
\qed

\medskip

\se{4. Proof of Proposition~2} 

In the rest of the paper, we prove our key estimate, Proposition~2, in several steps. 

\se{4.1. Reduction to a single particle problem} 
By the translation symmetry of the system,
we can focus on one box $\bo{\zz}$ at the origin without loss of generality, and  
we hereafter omit the subscript $\ve c$ 
in $\ha \r_{\ve c}$ and $\wh{\De\r}_{\ve c}$.
We define a Heisenberg operator
$\ha\r_H(t):=e^{i\hat{H} t}\,\ha{\r}\,e^{-i\hat{H} t}$,
and accordingly, $\wh{\De\r}_H(t)$. 
Note that the LHS of \eqref{eq-timeav-bd} satisfies 
\begin{align}
	& 
	\bigl [ \sw{\Psi(t)}{(\wh{\De\r})^2} \bigr ]_\ta
	= 
	\bigl [ \bra{\Psi(0)} (\wh{\De\r}_H(t))^2 \ket{\Psi(0)} \bigr ]_\tau
	\nonumber
	\\
	& 
	\quad = \bra{\Psi(0)} \, \bigl [ (\wh{\De\r}_H(t))^2 \bigr ]_\tau \ket{\Psi(0)} 
	. 
	\label{eq-LHSprop2-rewrite1}
\end{align}

We work in the momentum representation. 
The observable $\ha\r$ of  
\eqref{eq-rho} is rewritten in terms of 
$\da {\ha a}_\aa$ and ${\ha a}_\aa$ as
\begin{align}
  	\ha\r
  	&=
  	\f1{V}\sum_{\aa,\bb}
  	\ti w(\aa-\bb)\da {\ha a}_{\aa}{\ha a}_{\bb},
  	\label{eq-rho-in-a}
    \\
  	\ti w(\mm) 
  	&=
  	\f1{l^d}\sum_{\xx; \norminf \xx<\f l2}
  	e^{-2 \pi i\mm\cdot\xx / L}
  	=
    \prod_{\m=1}^d \ti w_1(m^\m), 
    \label{eq-tiw}
\end{align}
where
$\ti w_1(k):={\sin(\f{\pi lk}L)}\big/({l\sin\f{\pi k}L})$ for $k \neq 0$, 
and $\ti w_1(0) := 1$.
Note that $\ti w(\zz)=1$.

Since the Hamiltonian \eqref{eq-H-in-a} is diagonal 
in momentum representaiton,
we have 
$
  e^{it\ha H}\da{\ha a}_\aa =
  e^{itE_\aa} \da{\ha a}_\aa e^{it\ha H}$ and  
${\ha a}_\bb e^{-it\ha H}=e^{-itE_\bb}e^{-it\ha H}{\ha a}_\bb
$. 
Therefore, from 
\eqref{eq-rho-in-a}
we have
\begin{align}
	\ha\r_H(t)
  	=
  	\f1V\sum_{\mm} \ti w(\mm)\ha F_t(\mm), 
    \label{eq-rho-in-a-2}
\end{align}
where 
\begin{equation}
	\ha F_t(\mm):=  \sum_{\bb}
	e^{-it\ti E_{\bb,\mm}}
	\hat{a}^\dagger_{\bb+\ve m}\hat{a}_{\bb}, 
	\quad 
	\ti E_{\bb,\mm}:=E_\bb-E_{\bb+\mm} . 
	\label{eq-Etilde-def.11}
\end{equation} 
We observe that the 
time evolution operator
$e^{-it\ha H}$ in $\ha\r_H(t)$ has disappeared,
or ``pair-annihilated.'' 
As a result, the analysis of 
$\ha\r_H(t)$ has essentially been reduced to that of a
single particle with energy spectrum $\ti E$. 
We also have
\begin{align}
	\wh{\De\r}_H(t)=\ha\r_H(t)-\frac{N}{V}
	= \frac{1}{V}\sum_{\mm\ne\zz}\ti w(\mm)\ha F_t(\mm), 
	\label{eq-drt-by-mneq0}
\end{align}
because $\ti w(\zz)=1$ and
$\ha F_t(\zz)=\sum_{\bb}\da {\ha a}_\bb {\ha a}_\bb=N$ on $\HH$.

\se{Remarks}
(a) 
$\ti w_1$ satisfies the bound 
\footnote{%
  Eq.~\eqref{eq-w1-bound.11} holds for $m=0$
  because $\ti w_1(0) = 1$ by definition.  
  For $m \neq 0$, we note $\ti w_1(m)$ is even in $m$, and we assume $m >0$.  
  We note that $\ti w_1(m) = (\sin\f{\pi l m}L)/(l\sin\f{\pi m}L) 
  = (1/l) \sum_{x=-(l-1)/2}^{(l-1)/2} e^{- 2 \pi i m x/L}$. 
  Taking the absolute value of both sides proves $|\ti w_1(m)| \leq 1$.  
  Also, using $|\sin\f{\pi l m}L| \leq 1$ and $|\sin\f{\pi m}L | \geq \frac{2m}{L}$ 
  proves $|\ti w_1(m)| \leq L/(2 m l) \leq n/(2m)$. 
}%
\begin{equation}
	\abs{\ti w_1(m)}  
	\le 1\we \f n{2\abs{m}}. 
	\label{eq-w1-bound.11}
\end{equation}
where $a\we b:=\min\brac{a,b}$. 
This 
guarantees that only 
$m$'s with small $\norminf{\mm}$ are relevant in the sum over $\mm$ 
in \eqref{eq-drt-by-mneq0}, 
because $n$ is a constant independent of $L$ 
due to macroscopic nature of our observable $\ha\r$.

(b)
We have
$\tilde{E}_{\beta, \mm} = O(\norminf{\mm|}/L)$, as is
seen from \eqref{eq-tiE} below. 
Since $\norminf{\mm}$ is effectively small
as was just remarked, 
we expect
that the time average of $(\De\r)$ over a {\timescale} 
$\tau \gtrsim O(L)$ is small and that 
the equilibration {\timescale} is 
$O(L)$.

(c) However, to obtain precise information on the fraction of
nonequilibrium time 
as shown in Theorem~1$'$,
we must 
analyze not only $(\De\r)$ but $(\De\r)^2$. 
For this, we must deal with time 
averages of $e^{-it(\ti E_{\bb,\mm}-\ti E_{\cc,\ve n})}$,
which depends on \emph{two} energies, 
in \eqref{eq-FFexp.2} below. 
The following proof, based on the identity \eqref{eq-decomp-meas}, 
is so devised as to control time averages of this type efficiently.  
As a result,  we do not need minute
control of possible degeneracies 
of $\ti E_{\bb,\ve m}$'s \footnote{Note that nondegeneracy of  $\ti E_{\bb,\ve m}$ 
is equivalent to the so called ``non-resonance'' condition for $E_{\bb}$'s in our 
setting.}; some good bounds on the ``number of states'' 
$|\Omega_{\mm}(E)|$ [see \eqref{eq-Omega-def}]
are sufficient.

\se{4.2. Key formula for $\bigbrak{ (\wh{\De\r}_H(t))^2 }_\ta$}

Let us express  $\bigbrak{ (\wh{\De\r}_H(t))^2 }_\ta$ in a more tractable
form.
Since $\wh{\De\r}_H$ is Hermitian, 
we have from \eqref{eq-drt-by-mneq0}, 
\begin{align}
  	&{\bigbrak{ (\wh{\De\r}_H(t))^2 }_\ta}
  	=
    \frac{1}{V^2}
    \sum_{\mm,\ve n\ne\zz} \ti w(\mm)\ti w(\ve n)
    \bigbrak{ \ha F_t(\mm)\ha F_t(\ve n)^\dagger }_\ta, 
    \label{eq-to-FF}
  	\\
  	&\bigbrak{ \ha F_t(\mm)\ha F_t(\ve n)^\dagger }_\ta
  	=
  	\sum_{\bb,\cc}
  	\bigbrak{
  	e^{-it(\ti E_{\bb,\mm}-\ti E_{\cc,\ve n})}
  	}_\ta
    \da {\ha a}_{\bb+\mm} {\ha a}_{\bb} \da {\ha a}_{\cc} {\ha a}_{\cc+\ve n}.
    \label{eq-FFexp.2}
\end{align}
We now introduce our \emph{key identity}: for any $E'$ and $E''$, 
\begin{align}
	& 
	\bigbrak{
  	e^{-it( E' - E'')}
  	}_\ta
  	= \int_{-\infty}^{\infty} f_\tau(t) \, 
	e^{-it( E' - E'')}
	\nonumber \\
	& 
	= 
	\f\ta2
    \intinf \dd E\; 
    I\bigbrak{\nabs{E'-E }\le\f1\ta}\,
    I\bigbrak{\nabs{E-E'' } \le\f1\ta}, 
    \label{eq-decomp-meas}
\end{align}
where $I[\;\cdot\;]$ is the indicator function that takes value
1 if the argument is true and 0 otherwise.

\se{Remark}
This identity is a special case of a
well-known property of 
Fourier transform: if $f(t) = g(t)^2$, their Fourier transforms 
satisfy $\widehat{f} = \tx{{const.}} \, \hat{g} *  \hat{g}$, where $*$
denotes the convolution.
This general property enables us to 
take the sum over two energies $E'$ and $E''$ independently.
Our specific choice \eqref{eq-timeav-def} of $f_\tau$ leads to 
$\hat{g}(E) \propto  I[ |E| \le 1/\tau]$, 
which is most convenient for the present analysis. 

Applying identity \eqref{eq-decomp-meas} to \eqref{eq-FFexp.2}, we have  
\begin{align}
  	&\bigbrak{ \ha F_t(\mm)\ha F_t(\ve n)^\dagger }_\ta
    =
    \f\ta2
    \sum_{\bb,\cc}
    \intinf \dd E\;
  	I\bigbrak{\nabs{E-\ti E_{\bb,\mm}}\le\f1\ta}
    \nn
    &\qq \qq\times
 	I\bigbrak{\nabs{E-\ti E_{\cc,\ve n}}\le\f1\ta}\;
  	\da {\ha a}_{\bb+\mm} {\ha a}_{\bb} \da {\ha a}_{\cc} {\ha a}_{\cc+\ve n} 
    \nonumber \\
    & =
    \f\ta2
    \intinf \!\! \dd E\,
    \bigparen{ \sum_{\bb\in\Om_\mm(E)} \!\!
    \da {\ha a}_{\bb+\mm} {\ha a}_{\bb} }
    \bigparen{ \sum_{\cc\in\Om_{\ve n}(E)} \!\!
    \da {\ha a}_{\cc} {\ha a}_{\cc+\ve n} }, 
    \label{eq-FF}
\end{align}
where
\begin{equation}
	\Om_{\mm}(E):=\bigbrac{\bb; \;
  	\abs{\b^\m}<\f L2, 
  	\nabs{ E-\ti E_{\bb,\mm} }<\f1\ta} 
  	\label{eq-Omega-def}
\end{equation}
consists of the indices $\bb$ of ``energy levels
$\ti E_{\bb,\mm}$
near the energy $E$,'' with fixed $\mm$. 
Substituting \eqref{eq-FF} to \eqref{eq-to-FF}, we finally have 
\begin{align}
  	&
  	\bigbrak{ (\wh{\De\r}_H(t))^2 }_\ta
  	=
  	\f\ta2
    \intinf\dd E\;
    \ha G(E)\da{\ha G}(E),
    \label{eq-dr2t-final}
    \\
    &\hat{G}(E):=
    \f1V
    \sum_{\mm\ne\zz} \ti w(\mm)
    \sum_{\bb\in\Om_\mm(E)}
    \da {\ha a}_{\bb+\mm} {\ha a}_{\bb} . 
    \label{eq-dr2t-final-2}
\end{align}

\se{4.3. Our key estimate}

We use the following simple but useful inequality for 
the RHS of \eqref{eq-LHSprop2-rewrite1}: 
\begin{align}
  	\sw{\Psi(0)}
  	{\bigbrak{ (\wh{\De\r}_H(t))^2 }_\ta}
  	& \le
  	\bignormop{\bigbrak{ (\wh{\De\r}_H(t))^2 }_\ta}.
  	\label{eq-drt-le-drhto}
\end{align}
It thus suffices to bound the operator norm of \eqref{eq-dr2t-final}.
We have 
$\| \hat{G}(E) \|_{\rm op} \le \f1V \sum_{\mm\ne\zz} \abs{\ti w(\mm)} \abs{\Om_\mm(E)}
$ 
from \eqref{eq-dr2t-final-2} and 
$\nnormop{\da {\ha a}_{\bb+\mm} {\ha a}_{\bb}} =1$. 
Using the Schwarz inequality
$\f\ta{2}\intinf\dd E\;
\abs{\Om_\mm(E)}
\abs{\Om_{\ve n}(E)}
\le \sqrt{\II_\mm \II_{\ve n}}$,
where
\begin{equation}
	\II_\mm:=
	\f\ta{2}
	\intinf\dd E\;
	\abs{\Om_\mm(E)}^2
	, 
	\label{eq-IIdef.11}
\end{equation}
we obtain from \eqref{eq-dr2t-final},
\begin{align}
  	\bignormop{\bigbrak{ (\wh{\De\r}_H(t))^2 }_\ta}
  	&\le 
  	\bigl ( \f1{V}
  	\sum_{\mm\ne\zz}
  	\abs{ \ti w(\mm) }
    \sqrt{\II_\mm} \, \bigr )^2
    . 
 	\label{eq-drhto-ineq}
\end{align}
Our task has thus been reduced to estimating 
$\abs{\Om_\mm(E)}$, and its square integral $\II_\mm$ 
appearing in \eqref{eq-drhto-ineq}. 
Note that $\abs{\Om_\mm(E)}$ is a ``density of states (DOS),'' 
or rather ``number of states,''
of single particle energy eigenvalues $\tilde{E}$.

\se{4.4. Estimating $\abs{\Om_\mm(E)}$ for $d=1$}
Let us give a concrete estimate of 
$\abs{\Om_\mm(E)}$ for $d=1$. 
We write $\aa=\a$, $\mm=m$, etc.
Let $m\ne0$. 
Because \eqref{eq-H-in-a}
and \eqref{eq-Etilde-def.11} implies
\begin{align}
\ti E_{\b,m}
=C_m \sin\bigbrak{\f{2\pi}{L} \bigparen{\b+\f m2}},
\q C_m=4\sin\f{\pi m}{L}, 
\label{eq-tiE}
\end{align}
we can write 
$\abs{\Om_m(E)}=\bigabs{\om_m(\f E{C_m},\f1{C_m\ta})}$, 
where 
\begin{align}
  \om_m(x,\d):=\bigbrac{\b;\; \abs\b<\f L2,
  \Bigabs{x-\sin\bigbrak{\f{2\pi}{L}\bigparen{\b+\f m2}} }
  \le \d }.
  \label{eq-om}
\end{align}
It is obvious that $\abs{\om_m(x,\d)}=0$
for $\abs x> 1+\d$. 
For  $|x| \leq 1 + \d$, we have the following Lemma, 
whose proof is given in Appendix~C. 

\smallskip

\bfthm{Lemma 5} (Upper bound on the $\abs{\Om_\mm(E)}$ for $d=1$)
If $m\ne0$, $0<\d \le1/2$ and $\abs x\le 1+\d$, we have 
\begin{align}
  	\abs{\om_m(x,\d)}
  	\le
  	\f{L\d}{\sqrt{1-\{\abs x\we (1-\d)\}^2}}+2 . 
  	\label{eq-DOS}
\end{align}
$\abs{ \Om_m(E) }$ obeys this upper bound with 
$x=E/C_m$ and $\d =1/(C_m\ta)$. 

\smallskip

\se{4.5. Estimating $\II_\mm$}

We now estimate $\II_\mm$.
For $d=1$, a direct computation using Lemma~5 yields the following: 

\smallskip

\bfthm{Lemma 6} (Upper bound on $\II_m$ for $d=1$)
Suppose $d=1$,  $m\ne0$, $L>10^4$ and $\ta>2L$. Then we have 
\begin{align}
  	\II_m\le
  	L^2 \, \f{\log (\abs{C_m}\ta)}
  	{\abs{C_m}\ta} +4L.
\end{align}

\smallskip

For $d \geq 2$, we observe that
$\II_\mm$ are bounded by $\II_m$ of $d=1$:

\medskip

\bfthm{Lemma~7} (Upper bound on $\II_\mm$ for $d \geq 2$)
Suppose $d \geq 2$, $\mm\ne\zz$, $L>10^4$ and $\ta>2L$. Then, we have 
\begin{align}
	\II_\mm 
	\leq \f{V^2}{L^2} \II_{\norm \mm_\infty},
	\quad \norminf\mm = \max_{1\le\m\le d} \abs{m^\m}, 
	\label{eq-Imd>2-by-d=1}
\end{align}
where $\II_{\norm \mm_\infty}$ on the RHS denotes $\II$ of $d=1$.  

\smallskip

Proofs of these Lemmas are given in Appendices~D and E.

\se{4.6. Estimating 
  $\bigl \| \bigbrak{ (\wh{\De\r}_H(t))^2 }_\ta \bigr \|_{\rm op}$}

Using bounds on $\II_\mm$ given by Lemmas~6 and 7, 
we now bound the RHS of \eqref{eq-drhto-ineq}. 

\smallskip

\bfthm{Lemma~8}
For $L>10^4$ and $\ta>2L$, 
we have
\begin{align}
  	\f1V\sum_{\mm\ne\zz}\abs{\ti w(\mm)}\sqrt{\II_\mm}
  	\le
    K_{d}' \, n^d \, 
  	\d_{\f12}(\ta,L), 
\end{align}
where $K_d'>0$ depends only on $d$. 

\smallskip

The proof of Lemma~8 is given in Appendix~F.

\smallskip

We complete the proof of Proposition~2 by 
Eqs.~\eqref{eq-LHSprop2-rewrite1}, 
\eqref{eq-drt-le-drhto}, and 
\eqref{eq-drhto-ineq} and  Lemma~8, 
together with
$\nparen{\d_{\f12}(\ta,L)}^2\le 2\d_{1}(\ta,L)$.

\medskip

\se{5. Conclusion}

In this Letter, we considered 
a  class of translation-invariant free-fermion systems 
(including those with uniform nearest neighbor hopping) 
on a $d$-dimensional $L \times \cdots \times L$ hypercubic lattice. 
We proved that, starting from an \emph{arbitrary} pure initial state, 
the system equilibrates with respect to 
the coarse-grained density within a {\timescale} of order $L$.  
This scaling is optimal, 
since there exist initial states whose equilibration requires time of order $L$. 
Our result establishes $O(L)$ as the \emph{equilibration {\timescale}} for the macroscopic density. 
This will be the first time  that 
a {\timescale} of macroscopic equilibration has been rigorously established for
isolated physically natural quantum systems. 
Although our current result is restricted to free fermion systems (which are integrable), 
we hope our analysis will shed some light on 
our understanding of macroscopic equilibration in non-integrable quantum systems.

\medskip

\se{Acknowledgments}
It is a great pleasure to thank Hal Tasaki for numerous constructive
comments on an early version of the manuscript.  
We also thank anonymous referees for valuable comments. 
The present research was supported in part by JSPS Grants-in-Aid for 
Scientific Research, No. 18K03337 and No. 20K03772.

%

\appendix

\newpage

\newpage 

\section*{END MATTER}

\se{A. Proof of Lemma~3}
Since $\sum_{\ve c}(\wh{\De\r}_{\ve c}){}^2$
preserves the decomposition
$\HH=\HH\eqb\oplus\HH\nqb$ and 
$\sum_{\ve c}(\wh{\De\r}_{\ve c}){}^2\ge(\ep\ba\r)^2$
holds
on
$\HH\nqb$, we have, for any $\ket\Psi\in\HH$, 
\begin{align}
  \sw\Psi {\hat{P}\nqb}
  &\le
  \bra\Psi {\hat{P}\nqb}
  \f{\sum_{\ve c}(\wh{\De\r}_{\ve c}){}^2}{(\ep\ba\r)^2}
    {\hat{P}\nqb}\ket\Psi
    \nn
    &
    \le
      \bra\Psi \f{\sum_{\ve c}
      (\wh{\De\r}_{\ve c}){}^2}
      {(\ep\ba\r)^2}\ket\Psi. 
\end{align}

\se{B. Proof of Lemma~4}
Since \eqref{eq-timeav-def} implies
$f_\ta(t)\ge (\sin1)^2/(\pi\ta)$ for $|t|\leq \tau$,  
we have 
$
B\ge \brak{X(t)}_\ta
  \ge
  \bigbrak{X_1 I[X(t)>X_1]I[|t|\le \ta] }_\ta
  \ge
  X_1 \times
  (\sin1)^2/(\pi\ta) \times 2\ta \times
  \paren{\tx{fraction of time duration with \ma{X(t)>X_1}
      in \ma{[-\ta,\ta]}}}
$. 
That $2 (\sin 1)^2/\pi >1/3$
proves the Lemma. 
\qed

\medskip

\se{C. Proof of Lemma~5}
From \eqref{eq-om}, the problem is to find the number of integers
$y\in(-\pi,\pi]$
satisfying
$\sin\paren{\f{2\pi}L(y+\f m2)}\in[x-\d, x+\d]$,
which is twice the number of such integers 
$y\in(-\f\pi2,\f\pi2]$.
We assume $x\ge0$ without loss of generality
because
$\abs{\om_m(x,\d)}$ does not depend on the signature of $x$. 
When $1-\d<  x\le 1+\d$, we observe that 
$\abs{\om_m(x,\d)}\le\abs{\om_m(1-\d,\d)}$. 
When $0\le x\le 1-\d$, we have 
\begin{align}
  \abs{\om_m(x,\d)}
  \le 2\Bigl(
  \f {L}{2\pi}\paren{\Arcsin(x+\d)-\Arcsin(x-\d)}
  +1
  \Bigr)
\end{align}
{}(the range of $\Arcsin$ is $[-\pi/2,\pi/2]$). 
Therefore, Lemma~5 is proven 
(also from $2\sqrt2<\pi$) once the following inequality is proven 
for $0 < \delta \leq 1/2$ and $0\le x\le 1-\d$:
\begin{align}
  \Arcsin(x+\d)-\Arcsin(x-\d)
  \le \f{2\sqrt2 \d}{\sqrt{1-x^2}}.
  \label{eq-arcsin}
\end{align}
To prove \eqref{eq-arcsin},
let $\ep$ be the LHS.
Since $\ep$ is increasing in $x$ (easily seen by
differentiation with respect to $x$),
$\ep\le \ep|_{x=1-\d}= 
\Arcsin(1)-\Arcsin(1-2\d)<\pi/2$. 
Let $\tht:=\Arcsin(x+\d)$.
It is seen by taking squares of both hand sides
of \eqref{eq-arcsin} 
that 
\eqref{eq-arcsin} is equivalent to
  \begin{align}
    Y(\tht,\ep):=\f{8\paren{\sin\tht-\sin(\tht-\ep)}^2}
    {4-\paren{\sin\tht+\sin(\tht-\ep)}^2}-\ep{}^2
    \ge0. 
  \end{align}
$Y$ is non-increasing in $\tht$ because 
  \begin{align}
    \pder Y{\tht}
    =
    -\frac
    {128
    \bigparen{\sin\f{\ep}2}^4
    \sin (2 \tht-\ep)
    }
    {\bigl \{ 
    4-\paren{\sin\tht+\sin(\tht-\ep)}^2
    \bigr \}^2}
    \le0, 
  \end{align}
  where $x\ge0$ implies $\sin (2 \tht-\ep )\ge0$.
  Thus, 
  $Y(\tht,\ep)\ge Y(\pi/2,\ep)
  =\{8-3 \ep{} ^2-(8
  +\ep{} ^2) \cos \ep\}
  /\{3+\cos\ep\}$. 
    Because $\cos\ep\le1-\ep{}^2/2+\ep{}^4/24$, 
    (the numerator of $Y(\pi/2,\ep)$)
    $
     \ge \ep{} ^4 (2-\ep ) (2+\ep)/24. 
    $
    This is nonnegative because $\ep\in(0,\pi/2)$. 
    We have proven \eqref{eq-arcsin},
    and hence Lemma~5. 
\qed

\se{D. Proof of Lemma~6}

For $d=1$, $x=E/\abs{C_m}$ and $\d=1/(\abs{C_m}\ta)$,
we have 
\begin{align}
 & \II_m
  =
    \f1 {2\d}
    \int_{-1-\d}^{1+\d}
    \dd x\; 
  \abs{\om_m(x,\d)}^2
  \nn
  &\le
 \frac{1}{2\d}\int_{-1-\d}^{1+\d}\dd x\; 
    \Bigl ( 
    \f{(L\d)^2}{1-\{\abs x\we (1-\d)\}^2}+4\abs{\om_m(x,\d)}
    \Bigr ) .
    \label{eq-Im-bd}
\end{align}
In the last inequality, we used Lemma~5 
for the integrand 
and the fact that 
\footnote{Proof: the claimed 
inequality is equivalent to $(a-b)(a+b) \leq 2 a c$.  
This 
is trivially true for  $a < b$.
For $a \geq b$, we have 
$(a-b)(a+b) \leq c(a+b) \leq 2 a c$, because $a-b \leq c$.}
$a\le b+c$
for $a,b,c\ge0$ 
implies $a^2\le b^2+ 2ac$. 

The integral of $\abs{\om_m(x,\d)}$
equals the number $L$ of total possible states $\b$,
times the width $2\d$,
so that the contribution from the second integrand
in \eqref{eq-Im-bd} is exactly 
$4L$.

Performing the integral of the first integrand of \eqref{eq-Im-bd}, 
we see the RHS of \eqref{eq-Im-bd}
is at most  
\begin{align}
    {L^2 \delta}
    \Bigl (  
    \frac{1}{2} \log\f{2-\d}{\d}
    + \frac{2}{2-\delta} 
    \Bigr ) 
    +
    4L. 
    \label{eq-prf-lem-6.11}
\end{align}
What remains to be shown is that the quantity in the parenthesis of 
\eqref{eq-prf-lem-6.11} is bounded above by $\log(1/\delta)$,
or
\begin{equation}
	\log \delta + \log(2-\delta) + 4/(2-\delta) \leq 0 . 
	\label{eq-prf-Lemma6.21}
\end{equation} 
To show this,
we first note $\delta < 1/25$, 
because 
the definition \eqref{eq-tiE} 
of $C_m$ and the conditions $L>10^4$ and $\ta>2L$
imply that
$1/\d =\abs{C_m}\ta\ge 4\ta\sin\f\pi L
\ge 4\ta\f\pi L(1-10^{-5})>8\pi(1-10^{-5})>25$. 
The LHS of \eqref{eq-prf-Lemma6.21} is increasing in $\delta$, 
and is easily seen to be negative for $\delta = 1/25$. 
Thus, Lemma~6 has been proven.  
\qed

\medskip

\se{E. Proof of Lemma~7}

We reduce the case of $d\ge2$ to that of 
$d=1$. 
Fix $\mm$, and let $m$ be the component of $\mm$ with the largest magnitude
(take one if there are two or more).
We also decompose $\bb$ accordingly and write $\bb = (\b,\bb')$ 
(i.e., if $m$ is the $\m^{\rm th}$ component of $\mm$, $\b$ is the
 $\m^{\rm th}$ 
component of $\bb$).
Decomposing $\ti E_{\bb,\mm}=\ti E_{\b,m}+\ti E_{\bb',\mm'}$ in 
$
\Om_{\mm}(E):=\bigbrac{\bb; \;
  \abs{\b^\m}<\f L2, 
  \nabs{ E-\ti E_{\bb,\mm} }<\f1\ta} 
$, we observe that the DOS can be written as a sum of those 
in one dimension: 
\begin{equation}
  \abs{ \Om_\mm(E) }
  =
  \sum_{\bb'}
  \bigabs{ \Om_m(E-\ti E_{\bb',\mm'}) }
  . 
\end{equation}
Since $ab\le \f12(a^2+b^2)$, we have
\begin{align}
  &\abs{ \Om_\mm(E) }^2
  =
  \sum_{\bb',\cc'}
  \bigabs{ \Om_m(E-\ti E_{\bb',\mm'}) }
  \bigabs{ \Om_m(E-\ti E_{\cc',\mm'}) }
  \nn
  &\le
    \f12
    \sum_{\bb',\cc'}
    \bigparen{
  \bigabs{ \Om_m(E-\ti E_{\bb',\mm'}) }^2+
    \bigabs{ \Om_m(E-\ti E_{\cc',\mm'}) }^2
    }
  \nn
  &=
    {L^{d-1}}
    \sum_{\bb'}
  \bigabs{ \Om_m(E-\ti E_{\bb',\mm'}) }^2. 
\end{align}
Therefore, we
obtain 
\begin{align}
  \II_\mm
  &\le
    {L^{d-1}}
    \sum_{\bb'}
    \f\ta2
    \intinf \dd E\;
    \bigabs{ \Om_m(E-\ti E_{\bb',\mm'}) }^2
    \nn
    &\le
    L^{2d-2}\II_m
    =
    \f{V^2}{L^2}
      \II_{\norm \mm_\infty}, 
      \label{eq-Imm-bd}
\end{align}
where we have used $\II_m=\II_{\abs m}$. 
\qed

\medskip

\se{F. Proof of Lemma~8}

Because the proof is essentially the same for $d=1$ and for $d \geq 2$, we 
explain for $d \geq 2$.  
In the proof, $K_d\spp i$ ($i=1, 2, \ldots$) denotes 
a positive constant that depends only on $d$.

To evaluate
$\sum_{\mm\ne\zz}\abs{\ti w(\mm)}\sqrt{\II_\mm}$
in \eqref{eq-drhto-ineq} for $d \geq 2$,
we first take the sum over $\mm$ that has the same value of 
$\norminf\mm=:m$.
Since there are 
$2d$ possibilities for $m^\m=\pm m$,
the sum over $\mm\ne\zz$
is bounded by $2d$ times the value of 
the case of $m^1=+m$. 
Writing $\mm':=(m^2,...,m^d)$, and using Lemma~7 to 
bound  $\II_\mm$ in terms of 
$\II_m$ as $\II_\mm \leq (V^2/L^2) \, \II_m$, 
we have 
\begin{align}
  &
  \f1V\sum_{\mm\ne\zz}\abs{\ti w(\mm)}\sqrt{\II_\mm}
    \nn
  &\le \f{2d}L
    \sum_{1\le m<L/2}
    \abs{\ti w_1(m)}
    \sqrt{\II_m}
    \sum_{\norminf{\mm'}\le m}
    \abs{\ti w(\mm')}
    \nn
  &= \f{2d}L
    \sum_{1\le m<L/2}
    \abs{\ti w_1(m)}
    \sqrt{\II_m} \, 
    \Bigparen{ \sum_{k=-m}^m \abs{\ti w_1(k)} }^{d-1}.
    \label{eq-sum-sep}
\end{align}

Let us estimate each part of the sum.
First, using \eqref{eq-w1-bound.11} 
and
splitting the sum over $m$, 
we
have
\footnote{Since the summand is decreasing in $|k|$, we can use the 
integral to bound the sum rigorously.  Similar remarks apply to 
subsequent similar treatments.}
\begin{align}
  &\sum_{k=-m}^m
    \abs{\ti w_1(k)}
    \le
    1+2\Bigceil{\f n2}+
    2\sum_{k=\ceil{\f n2}+1}^m \f n{2k}
    \nn
    &
    \leq 
    1+2\Bigceil{\f n2} + n \int_{\ceil{\f n2}}^m\f{\dd x}x
      \le n+2+n\log m.
    \label{eq-sum-w}
\end{align}
This can be further simplified 
by
$m \leq L/2$  and $L>10^4$, 
\begin{align}
  \sum_{k=-m}^m
  \abs{\ti w_1(k)}
  &\leq
    2 n \log L.
    \label{eq-sum-w-simple}
\end{align}
Second, because $\abs{C_m} \geq 8|m|/L$ holds from \eqref{eq-tiE}, 
Lemma~6 implies 
\begin{align}
	\II_m \leq 
	L^2 \, \frac{\log (\frac{8 |m| \tau}{L})}{\frac{8 |m| \tau}{L}} 
	+ 4L. 
\end{align}
Since $\sqrt{a+b}\le\sqrt a+\sqrt b$ 
for $a,b\ge0$, 
this implies 
\begin{align}
  \sqrt{\II_m}
  	&\le
 	\frac{L \{\log (\frac{8 \tau}{L})\}^{1/2} 
	+ L \{\log m\}^{1/2}}{\bigl (\frac{8 m \tau}{L} \bigr )^{1/2}}
  + 2 \sqrt{L} . 
    \label{eq-sqrt-IIm}
\end{align}

We now plug
these 
into \eqref{eq-sum-sep}.
From \eqref{eq-sum-w-simple}, 
the contribution to \eqref{eq-sum-sep} from 
the second term of \eqref{eq-sqrt-IIm}
is bounded above by 
\begin{align}
  \frac{2d}{L}
  (2 n \log L)^{d-1} \times 2 \sqrt{L}
    \sum_{1\le m<L/2}
    \abs{\ti w_1(m)} . 
	\label{eq-2sqrtL.11}
\end{align}
The sum
in \eqref{eq-2sqrtL.11} 
is bounded as in
\eqref{eq-sum-w-simple},  so that 
\eqref{eq-2sqrtL.11} is bounded by 
\begin{align}
	K^{(1)}_d (n \log L)^d/\sqrt{L} .
	\label{eq-2sqrtL.13}
\end{align}

To bound the contribution  to \eqref{eq-sum-sep} from the first term of \eqref{eq-sqrt-IIm}, 
we first simplify a power of the RHS of \eqref{eq-sum-w} as 
$(n+2+n\log m)^{d-1}\le
K^{(2)}_d n^{d-1} \{1+(\log m)^{d-1} \}$,
and then 
bound the product of $\{1+(\log m)^{d-1} \}$ and  
the numerator of the first term of \eqref{eq-sqrt-IIm} 
as 
\begin{align}
	& 
	\{1+(\log m)^{d-1} \}
	\bigl \{ L \{\log (\tfrac{8 \tau}{L})\}^{1/2} 
	+ L \{\log m\}^{1/2} \bigr \} \, 
	\nonumber \\
	& 
	\leq K^{(3)}_d L \{\log (\tfrac{8 \tau}{L})\}^{1/2} \, m^{1/4} 
	+ K^{(4)}_d L \, m^{1/4} 
	\nonumber \\
	& 
	\leq 
	K^{(5)}_d \, L \,  \{\log (\tfrac{8 \tau}{L})\}^{1/2} \, m^{1/4} 
	.
\end{align}
In the above, we first expanded the product, then  bounded 
various powers of $\log m$ by 
$K^{(i)}_d m^{1/4}$ with appropriate choices of $K^{(i)}_d$, and 
finally used the fact that  $\log(\frac{8 \tau}{L}) \geq 1$, 
to combine the two terms into one. 
Using also \eqref{eq-w1-bound.11}, we bound 
the contribution to \eqref{eq-sum-sep} 
from the first term of \eqref{eq-sqrt-IIm}   by 
\begin{align}
	K^{(6)}_d n^{d-1} \Bigl ( \frac{\log (\frac{8 \tau}{L})}{\frac{8 \tau}{L}}
  	\Bigr )^{1/2} 
	\sum_{1\le m<L/2}
	\Bigl ( 1 \wedge \frac{n}{2m} \Bigr ) \frac{1}{m^{1/4}}
	.
	\label{eq-log-contri.11}
\end{align}
The sum over $m$ is handled as was done in \eqref{eq-sum-w}, 
and is bounded by $K^{(7)}_d n$.  As a result, 
the contribution from the first term of \eqref{eq-sqrt-IIm} 
is bounded by 
\begin{equation}
  K^{(8)}_d n^{d} \Bigl ( \frac{\log (\frac{8 \tau}{L})}{\frac{8 \tau}{L}}
  \Bigr )^{1/2}
  \le
  K^{(8)}_d n^{d} \Bigl ( \frac{\log (\frac{\tau}{L})}{\frac{\tau}{L}}
  \Bigr )^{1/2} , 
  \label{eq-eq-log-contri.31}
\end{equation}
since $\log x/x$ is decreasing in $x>1$. 

Combining \eqref{eq-2sqrtL.13} and  \eqref{eq-eq-log-contri.31}
proves 
Lemma~8. 
\qed

\end{document}